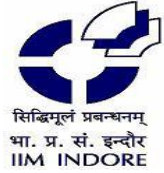

सिद्धिमूल प्रबन्धम्,
भा. प्र. सं. इन्दौर
IIM INDORE

# Enhancing Social Media Personalization: Dynamic User Profile Embeddings and Multimodal Contextual Analysis Using Transformer Models

Author: Pranav Vacharajani, EPGDPM & AI, Batch – 2, Indian Institute of Management, Indore
Mentor: Prof. Pritam Ranjan, Indian Institute of Management, Indore

**Abstract**—This study investigated the impact of dynamic user profile embedding on personalized context-aware experiences on social networks. A comparative analysis of multilingual and English transformer models was performed by analyzing a dataset of over twenty million data points It involved analyzing a wide range of metrics and performance indicators to compare dynamic profile embeddings vs. non-embeddings. effectively static profile embeddings. A comparative study was conducted using degradation functions. Extensive testing and extensive research confirmed that dynamic embedding successfully tracks users' changing tastes and preferences, provides more accurate recommendations and user engagement Thus these results are important for social media platforms that aim to improve user experience through relevant features and sophisticated recommendation engines.

◆

## 1 KEYWORDS: DYNAMIC EMBEDDINGS, DECAY FUNCTIONS, TRANSFORMER MODELS, USER PROFILE PERSONALIZATION, SOCIAL MEDIA, MULTIMODAL DATA, RECOMMENDATION SYSTEMS, COSINE SIMILARITY.

## 2 INTRODUCTION:

2.1. **Background information and context:** In the rapidly evolving digital age , social media platforms are constantly looking for ways to increase user engagement and satisfaction.. Personalized content recommendations are central to this pursuit,[1]. requiring the development of sophisticated analytical techniques that can adapt to user behavior and preferences.. Transformer models known for their effectiveness in a variety of natural language processing tasks offer promising approaches has improved recommendation processes through embeddings of dynamic user profiles.[2]

2.2. **Problem Statement:** Traditional static user profile embeddings often fail to capture the temporal evolution of user interactions.. This results in inadequate recommendations. This limitation presents a challenge in maintaining user engagement, as recommendations do not reflect recent changes in user preference or behavior. [3]

2.3. **Purpose and Significance of the Study:** The main purpose of this study is to investigate the effectiveness of dynamic user profile embedding, using different decay functions, to accurately track and display the changes in user interest over time. The objectives of this study are:

• Evaluate how different the decay functions influence the relevance and accuracy of recommendations.
• Compare the performance of multiple transformer-based models in handling dynamic embeddings..
• Explore the implication of these finding for enhancing personalized user experiences on SM platforms.

## 3 LITERATURE REVIEW:

### 3.1 Dynamic User Profile Embeddings

In order to deal with the dynamic nature of user interests and behaviours on the social media platforms like Twitter, Facebook, Instagram.. researchers have looked into dynamic user profile embeddings for better personalised recommendation system. In order to capture the temporal changes in user profiles, Liang et al. (2018) [4] suggested dynamic user and word embeddings, with the aim to capture the temporal changes in user profiles and activities. This can be instrumental in enhancing personalization.

Similar to this, Kerin et al. (2019) [5] highlighted the significance of temporal word embeddings for creating temporal user profiles, with the main driving force for their research being the dynamism present in user profiles. Furthermore, Zheng et al. (2018)[6] created a technique called user profile embedding that allows to capture similarities among users in social networks through user profile embedding (Zheng et al., 2018)[6]

### 3.2 Multimodal Contextual Analysis Using Transformer Models

Transformer models [7] ,especially those based on self-attention mechanisms, have revolutionised natural lan-

• Authors, Pranav Vacharajani, Pritam Ranjan are with Indian Institute of Management Indore, Indian Institute of Management Indore

guage processing and have been applied for recommendation systems. They are perfect for analysing noisy and unstructured social media data because of their capacity to handle sequential data and capture long-range dependencies. Wu et al.'s survey from 2023 [8] examines the different ways that LLMs can be used in recommendation systems and emphasises how well they work to produce personalised content using advanced pattern recognition and predictive analytics.

A major area of recent study has been the promise of Transformer models in multimodal learning environments. The main elements and difficulties of using Transformers to integrate multimodal data were covered by Xu et al. (2023) (Xu et al., 2023) [9]. To analyse unaligned multimodal language sequences, Tsai et al. ((Tsai et al., 2019).)[9] presented the Multimodal Transformer (MulT), which makes use of directional paired cross-modal attention mechanisms. Bartolomeu et al. (2022) [10] developed a context-enriched Multimodal Transformer model that enhances textual and visual contexts to better align images with news content, thus expanding on this idea (Bartolomeu et al., 2022).[10]

### 3.3 Social Media Recommender Systems Approaches

Recommender systems (RS) are (Eirinaki et al., 2018) [11] used by a variety of social media platforms to improve user experience through engagement and content personalisation. Anandhan et al. (Alamdari et al., 2020)[12] examined methods in social media RS, emphasising the use of collaborative filtering and content-based filtering, frequently in hybrid forms, and concentrating on context-awareness, scalability issues, and decision analysis. In their 2018 study, Eirinaki et al. (Eirinaki et al., 2018) [11] concentrated on the handling of user-generated content and volatile social interactions by RS in large-scale networks, highlighting the necessity of scalable solutions that can adjust to changing user preferences. Zhou et al. (Zhou et al., 2012) [13] talked about the development of personalised RS in social networking, including its history, technological advances, and social settings improvements.

### 3.4 E-commerce Recommender Systems Approaches

Recommendation systems (RS) improve the e-commerce shopping experience and boost sales. An overview of RS techniques, such as collaborative filtering, content-based filtering, knowledge-based systems, and hybrid systems, were given by Alamdari et al. (Alamdari et al., 2020) [12] . Sivapalan et al. (Sivapalan et al., 2014) [14] investigated how RS increases the effectiveness of e-commerce by recognising patterns in user behaviour and making product recommendations in line with those patterns. Furthermore, Schafer et al. (Schafer et al., 1999) [15] . described how RS helps e-commerce sites by making product recommendations based on user ratings, browsing history, and past purchases.

### 3.5 Main Approaches for Recommender Systems

In the digital world, recommender systems are essential because they direct user experiences by making relevant suggestions for products, content, and social interactions based on the interests and actions of the user. These systems use a range of approaches, each with special advantages and modifications for distinct platforms, such as social media (SM) websites and e-commerce platforms. The main approaches and their applications across notable platforms are mentioned below. [3], [12]

#### 3.5.1 Collaborative Filtering (CF)

In recommender systems, one of the most popular methods is collaborative filtering. It bases its recommendations on user preferences as a whole, presuming that users who have previously agreed will do so in the future.

• User-Based CF: This technique is used by eBay and Amazon to make product recommendations by comparing users based on past purchases. [16]

• Item-Based CF: YouTube uses this method to make video recommendations by analyzing the similarities in user interactions with different content. [17]

#### 3.5.2 Content-Based Filtering (CBF)

Based on feature similarity, content-based filtering suggests products that are similar to those a user has previously favoured. One of the best examples of this is Netflix, which provides recommendations to users based on the genres, actors, and directors of films and TV shows they have already seen[18].

#### 3.5.3 Hybrid Approaches

Hybrid systems combine content-based filtering and collaborative filtering to avoid certain limitations of each approach. Spotify employs this technique to provide personalised music recommendations by combining user behaviour (collaborative data) and song audio features (content-based data).[16]

#### 3.5.4 Knowledge-Based Systems

These systems recommend products based on explicit knowledge about the item assortment, user preferences, and recommendation criteria. This approach is often used in situations where items are not purchased frequently, such as on real estate or job sites like LinkedIn[19] .

#### 3.5.5 Demographic-Based Recommendations

This approach makes product recommendations based on user demographic data. Facebook and Instagram match recommendations with users' age, gender, location, and interests by using demographic data to target advertising and content[20].

#### 3.5.6 Utility-Based Systems

Products are recommended by utility-based recommender systems based on an assessment of their value to a specific user. This approach is used by Amazon's Alexa to make product recommendations based on the usefulness or practicality determined from user interactions and queries[21].

#### 3.5.7 Session-Based Recommendations

These systems provide recommendations based on what was done during a session. This method is common in e-commerce settings for brief sessions; it can be observed on platforms such as Alibaba, where real-time product recommendations are provided based on user interactions within a session[22] .

### 3.5.8    Deep Learning Approaches

Deep learning techniques have have gained popularity because of their capacity to represent intricate nonlinear relationships in data. Convolutional neural networks (CNNs) are used by Pinterest to evaluate visual content and suggest related pins. In order to recommend apps in its Play store, Google leverages deep neural networks in conjunction with user data and app features[23] .

Industry Applications

- Facebook and Instagram: Leverage a mixture of demographic-based and utility-based systems to optimize ad placements and content feeds[24].

- Amazon: Primarily uses collaborative filtering for product recommendations but also incorporates content-based methods and utility systems, especially in its voice-activated tool, Alexa[25] .

- eBay: Uses a hybrid strategy that combines elements of item-based collaborative filtering and user-based filtering to customise product recommendations based on user browsing and purchase history [26]

### 3.6    Identification of Gaps in the Literature

The emphasis on text in current models often obscures the dynamic user behaviours and multimodal data that are crucial for comprehending user preferences. (Recommender Systems Handbook.pdf; Social Media Recommendation Algorithms Tech Primer.pdf) [16], [27]. Furthermore, a lot of models don't make full use of the real-time data processing that's required to capture ephemeral social media trends. The embedding which is very important aspect is needs to be addressed.

### 3.7    How the Current Research Addresses These Gaps

This study uses transformer models to create a model that combines multimodal data with dynamic user profile embeddings. By updating user profiles dynamically in real-time, it overcomes the drawbacks of static profiling. The model seeks to improve contextual relationship understanding by utilising transformer technology, which will allow for more precise user preference predictions and greatly improved personalised recommendations.

## 4    METHODOLOGY

### 4.1    Research Design and Methods

The study adopted a quantitative research design to systematically check the effectiveness of dynamic user profile embeddings as compared to static ones.. Various transformer models, [28] such as multilingual and English-specific ones, have been analyzed the use of special decay capabilities [29] to understand their effect on the accuracy and relevancy of content recommendations on social networks.

### 4.2    Tools and Techniques for Data Analysis

The analysis was conducted using several advanced data processing and machine learning tools:

• Python: Used for scripting and automating the data analysis process, including data cleaning and transformation tasks.

• Pandas and NumPy: Employed for efficient data manipulation and numerical analysis.

• Scikit-learn: Utilized for implementing machine learning models and conducting statistical tests to compare the effectiveness of different embedding techniques.

• TensorFlow and PyTorch: Applied for training and evaluating deep learning models, particularly for developing and testing the multimodal profile embeddings.

• Sentence Transformers Library: Used to generate and manipulate embeddings from the transformer models.

• GPU : Kaggle's NVIDIA TESLA P100 GPU was used as the process required high computing power.

• GitHub: GitHub repositories were used for keeping the codes [30]

## 5    DATA COLLECTION:

### 5.1    Identifying Influential Personalities:

To start with data collection exercise, we identified the top **100 most followed personalities** on the Twitter (now its X), from various fields & regions globally, reflecting diverse interests. This selection aimed to cover a broad spectrum of user engagement.

### 5.2    User Data Compilation:

Utilizing the Twitter API, we extracted data from these 100 influential users along with their 1,000 most active followers each, culminating in a dataset comprising **100,100 individual user profiles.**.

### 5.3    Tweet Data Acquisition:

We proceeded to gather timeline data for these users, resulting in approx **20 million data points**. Post-processing, which included the removal of duplicates, refined the dataset to the final count of **14,765,661 (14M)** data points.

### 5.4    User Activity Data:

To analyze user interactions such as... likes, Quo Tweets, and ReTweets, we collected user activity data. For evaluative purposes, we randomly selected 30,000 instances from this dataset. The total of all liked tweets from these selected instances amounted to **2,107,054 (2M)** data points.

This extensive data collection phase spanned approximately three weeks, constrained by the rate limits by the Twitter API.[31]

### 5.5    Concatenation of the data

After downloading the Twitter activity data, the next step involves processing the data using a comprehensive approach that includes concatenation, parsing, cleaning, and storage. Multiple JSON and CSV files containing Twitter activity timelines and user tweets are imported from various sources. Data from CSV files are read into Pandas DataFrames, with columns parsed and relevant fields extracted, followed by filtering to include specific information such as timestamps, user names, user descriptions, and tweet texts. Parsed data from individual files are appended into a list of DataFrames, which are then concatenated into a single DataFrame to create a unified dataset. Memory management techniques, such as garbage collection are utilized to handle large datasets efficiently, and dtype warnings are managed to ensure smooth data import despite mixed data types. The final concatenated DataFrame is saved to a CSV file for further analysis, providing a robust foundation for subsequent analysis and modeling efforts, such as creating dynamic user profile embeddings.

[link of code on Github : https://github.com/Pranav-V-27/DE_Project ]

# 6 Algorithm Flow Chart

A dynamic profile embedding method for social media data is shown in the flow chart. The first step in the process is gathering user information from Twitter (Now X) , such as their bio and tweets. Hybrid tweet embeddings are produced by processing each tweet along with the user bio through an encoding model. A decay function is applied to these embeddings to adjust their weights based on their temporal relevance and recentness. A dynamic profile embedding that reflects the user's changing interests and preferences is the result. The most advantageous aspect of this strategy is its capacity to dynamically update the user profile with the most recent data, guaranteeing that recommendations and personalisations are current and pertinent. By utilising the most recent user activity data, this dynamic feature improves the accuracy of predictions pertaining to user engagement, such as predicting likes or follows.

# 7 Time Decay Functions and Dynamic Embeddings Analysis

## 7.1 Overview

Time Decay Functions and Dynamic Embeddings Analysis leverages various pre-trained models from the Sentence Transformers library (https://sbert.net/docs/sentence_transformer/pretrained_models.html ) to create embeddings from the textual data and than applies different time decay functions on these embeddings. Main goal is to know that how time influences the representation of textual data in a dynamic context.

With the application of these time decay functions, the code aims to simulate real-world scenarios where the relevance of information changes over time..

[link of code on Github : https://github.com/Pranav-V-27/DE_Project ]

## 7.2 Approach

The approach to analyzing the impact of various time decay functions, various models and similarities and diversities on dynamic embeddings covers many steps.. These steps are explained as below:

### 7.2.1 Data Loading and Data Preparation

Data Acquisition and Preprocessing: Data is been collected, now its necessary to convert and preprocess these data in the format which can be analysed.

### 7.2.2 Model Selection and Embedding Calculation

Model Selection: Based on the requirement, and existing resources, computing power, select appropriate pre-trained models. requirements.

### 7.2.3 Application of Time Decay Functions

Decay Function Selection: Analyse the application and relevance of various decay functions and choose a variety of decay functions (e.g., exponential, logarithmic) to model to catch the relevance of data over time.

### 7.2.4 Integration and Assessment of Similarity Measures

Similarity Metric Integration: Employ Basic, Cosine, and Cos-time similarities to evaluate changes in embeddings.

## 7.3 Uniqueness of the Approach

7.3.1. Diverse Model Usage: Utilisation of various pre-trained models enables a comprehensive evaluation across different data types and model architectures, illustrating how each model captures and represents data dynamics.

7.3.2. Variety of Time Decay Functions: By experimenting with multiple decay functions, the analysis gains depth in understanding the differential impact of time on data relevance, catering to diverse application needs where the freshness of data varies in importance.

7.3.3. Integration of Multiple Similarity Measures: The approach not only applies traditional similarity metrics but also explores their integration into the decay functions. Incorporating a range of similarity measures, specifically Basic, Cosine, and Cos-time similarities, into the decay functions introduces a novel way to adjust embeddings based on temporal and semantic changes. This varied approach not only assesses the immediate similarity but also considers the evolution of these similarities over time, offering a multidimensional analysis of data dynamics.

## 7.4 Model Selection and Embedding Calculation

Following table gives and overview of the existing selective sentence transformer models.

(https://sbert.net/docs/sentence_transformer/pretrained_models.html )

Four different pre-trained models from the Sentence Transformers library are selected for our research:

- **all-MiniLM-L6-v2:** Small and fast. (https://huggingface.co/sentence-transformers/all-MiniLM-L6-v2 )
- **distiluse-base-multilingual-cased-v2:** Multilingual. (https://huggingface.co/sentence-transformers/distiluse-base-multilingual-cased-v2 )
- **all-mpnet-base-v2:** Large with extensive knowledge.( https://huggingface.co/sentence-transformers/all-mpnet-base-v2)
- **jinaai /jina-embeddings-v2-base-en**: State-of-the-art (SOTA). (https://huggingface.co/jinaai/jina-embeddings-v2-base-en)

**Algorithm**: Generate Text Embeddings with Multiple Models

Input: DataFrame columns 'bio' and 'tweettext'
Output: Embeddings files for each model
Procedure:

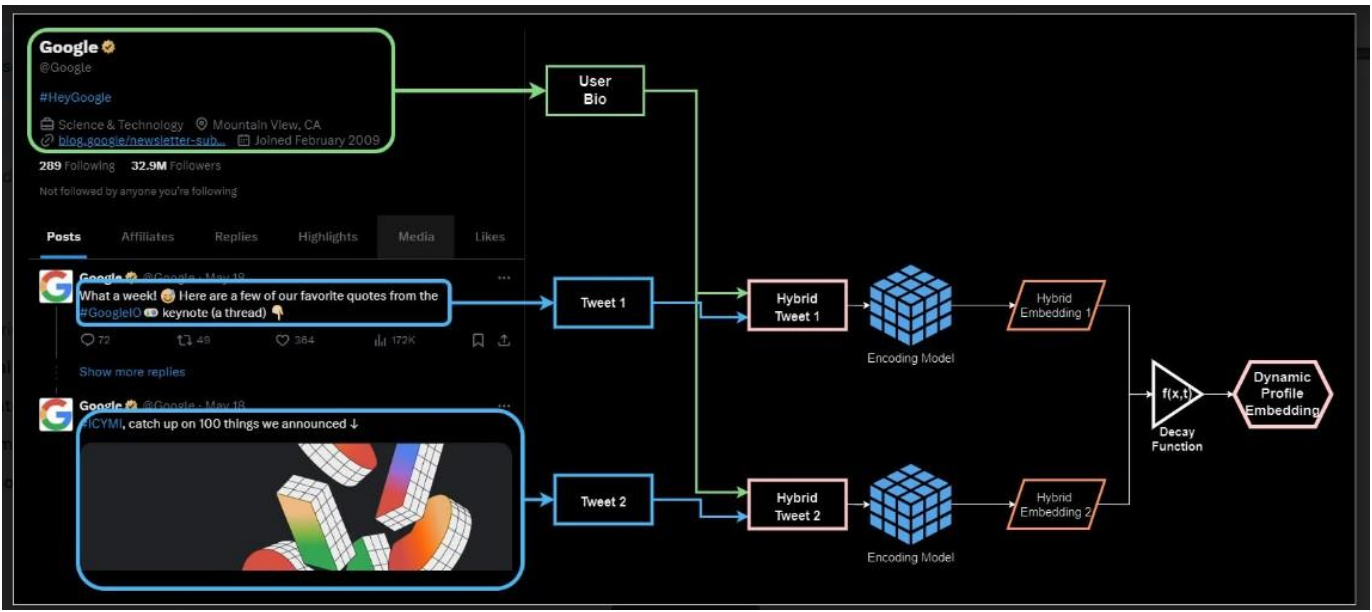

Fig. 1. 1 Flow chart

| Model Name | Performance Sentence Embeddings (14 Datasets) ⓘ | Performance Semantic Search (6 Datasets) ⓘ | ⇅ Avg. Performance ⓘ | Speed ⓘ | Model Size ⓘ |
|---|---|---|---|---|---|
| all-mpnet-base-v2 ⓘ | 69.57 | 57.02 | 63.30 | 2800 | 420 MB |
| multi-qa-mpnet-base-dot-v1 ⓘ | 66.76 | 57.60 | 62.18 | 2800 | 420 MB |
| all-distilroberta-v1 ⓘ | 68.73 | 50.94 | 59.84 | 4000 | 290 MB |
| all-MiniLM-L12-v2 ⓘ | 68.70 | 50.82 | 59.76 | 7500 | 120 MB |
| multi-qa-distilbert-cos-v1 ⓘ | 65.98 | 52.83 | 59.41 | 4000 | 250 MB |
| all-MiniLM-L6-v2 ⓘ | 68.06 | 49.54 | 58.80 | 14200 | 80 MB |
| multi-qa-MiniLM-L6-cos-v1 ⓘ | 64.33 | 51.83 | 58.08 | 14200 | 80 MB |
| paraphrase-multilingual-mpnet-base-v2 ⓘ | 65.83 | 41.68 | 53.75 | 2500 | 970 MB |
| paraphrase-albert-small-v2 ⓘ | 64.46 | 40.04 | 52.25 | 5000 | 43 MB |
| paraphrase-multilingual-MiniLM-L12-v2 ⓘ | 64.25 | 39.19 | 51.72 | 7500 | 420 MB |
| paraphrase-MiniLM-L3-v2 ⓘ | 62.29 | 39.19 | 50.74 | 19000 | 61 MB |
| distiluse-base-multilingual-cased-v1 ⓘ | 61.30 | 29.87 | 45.59 | 4000 | 480 MB |
| distiluse-base-multilingual-cased-v2 ⓘ | 60.18 | 27.35 | 43.77 | 4000 | 480 MB |

Fig. 2. Overview of the sentence transformer models

Initialize Models:
  model1 <- Load SentenceTransformer('sentence-transformers/all-MiniLM-L6-v2') with CUDA
  model2 <- Load SentenceTransformer('sentence-transformers/distiluse-base-multilingual-cased-v2') with CUDA
  model3 <- Load SentenceTransformer('sentence-transformers/all-mpnet-base-v2') with CUDA
  model4 <- Load SentenceTransformer('jinaai/jina-embeddings-v2-base-en') with CUDA
  Prepare Text Data:
  text <- Concatenate DataFrame columns 'bio' and 'tweet-text' into a list
  Compute Embeddings:
  main_embeddings1 <- Encode text with model1 using CUDA
  main_embeddings2 <- Encode text with model2 using CUDA
  main_embeddings3 <- Encode text with model3 using CUDA
  main_embeddings4 <- Encode text with model4 using CUDA
  Save Embeddings:
  Save main_embeddings1 to 'miniml6v2.npy'
  Save main_embeddings2 to 'distiluse-base-multilingual.npy'
  Save main_embeddings3 to 'all-mpnet-base-v2.npy'
  Save main_embeddings4 to 'jina-v2-en.npy'
  End Procedure

The decay functions play a crucial role in dynamic embeddings by assigning weights to embeddings based on their temporal distance. Here are the key decay functions :

Various time decay functions are implemented to adjust embeddings over time. Each function takes a timeline, embeddings, and a decay constant 'k' as inputs, and returns adjusted embeddings.

  Algorithm: Apply Time Decay Functions to Embeddings
  Input:
  timeline: Array of time points
  embeddings: Matrix of embeddings corresponding to each time point
  k: Decay parameter
  Output:
  Adjusted embeddings after applying decay
  Procedure:
  Define exponential_decay(timeline, embeddings, k):
  a <- Create an array from 1 to the length of the timeline
  decay <- Calculate $e^{-k*a}$
  Result <- Multiply each decay value with corresponding embeddings
  Return result
  Define inverse_linear_decay(timeline, embeddings, k):
  a <- Create an array from 1 to the length of the timeline
  decay <- Calculate $1 / (1 + k * a)$
  Result <- Multiply each decay value with corresponding embeddings
  Return result
  Define inverse_square_root_decay(timeline, embeddings, k):
  a <- Create an array from 1 to the length of the timeline
  decay <- Calculate $1 / \sqrt{1 + k * a}$
  Result <- Multiply each decay value with corresponding embeddings
  Return result
  Define hyperbolic_decay(timeline, embeddings, k):
  a <- Create an array from 1 to the length of the timeline
  decay <- Calculate $1 / (1 + k * a^2)$
  Result <- Multiply each decay value with corresponding embeddings
  Return result
  Define logarithmic_decay(timeline, embeddings, k):
  a <- Create an array from 1 to the length of the timeline
  decay <- Calculate $1 / \log(1 + k * a + 1)$
  Result <- Multiply each decay value with corresponding embeddings
  Return result
  Define gaussian_decay(timeline, embeddings, k):
  a <- Create an array from 1 to the length of the timeline
  decay <- Calculate $e^{-k*a^2}$
  Result <- Multiply each decay value with corresponding embeddings
  Return result
  End Procedure

7.6. **Similarities Calculation**: To measure the effectiveness of the time decay functions, specifically Basic, Cosine, and Cos-time similarities between consecutive pairs of embeddings is computed.

  Algorithm: Apply Time and Cosine-Similarity Adjusted Decay Functions to Embeddings
  Input:
  timeline: Array of time points (can be datetime objects)
  embeddings: Matrix of embeddings corresponding to each time point
  k: Decay parameter
  Output:
  Adjusted embeddings after applying decay
  Procedure:
  Define compute_cosine_similarity(embeddings):
  - Compute pairwise cosine similarity of the embeddings
  - Extract similarities for consecutive embedding pairs
  - Return the consecutive pair similarities
  Define compute_time_differences(timeline):
  - Calculate the time differences between consecutive timestamps in the timeline
  - Convert time differences to seconds if using datetime objects
  - Return the time differences
  Define decay_functions(timeline, embeddings, k) for each decay type:
  Fetch delta_t from compute_time_differences(timeline)
  Fetch cos_sim from compute_cosine_similarity(embeddings)
  For each decay function:
  Exponential Decay:
  Compute decay as $\exp(-k * a * delta\_t / cos\_sim)$
  Inverse Linear Decay:
  Compute decay as $1 / (1 + k * a * delta\_t / cos\_sim)$
  Inverse Square Root Decay:
  Compute decay as $1 / \sqrt{1 + k * a * delta\_t / cos\_sim}$
  Hyperbolic Decay:
  Compute decay as $1 / (1 + (k * a^2 * delta\_t) / cos\_sim)$
  Logarithmic Decay:

| Name | Significance | Formula (without $\Delta v$) | Formula (with $\Delta v$) | Formula (with $\Delta v$ and $\Delta t$) |
|---|---|---|---|---|
| Exponential Decay | Decays rapidly at first and then slows down. | $f(t) = e^{-\lambda t}$ | $f(t, \mathbf{v}_t, \mathbf{v}_{t-1}) = e^{-\lambda t} \cdot \Delta v$ | $f(t, \mathbf{v}_t, \mathbf{v}_{t-1}) = e^{-\lambda t \cdot \Delta t / \Delta v}$ |
| Inverse Linear Decay | Decays at a constant rate. | $f(t) = \frac{1}{1+\lambda t}$ | $f(t, \mathbf{v}_t, \mathbf{v}_{t-1}) = \frac{\Delta v}{1+\lambda t}$ | $f(t, \mathbf{v}_t, \mathbf{v}_{t-1}) = \frac{1}{1+\lambda t \cdot \Delta t / \Delta v}$ |
| Inverse Square Root Decay | Decays slowly, providing a more gradual reduction. | $f(t) = \frac{1}{\sqrt{1+\lambda t}}$ | $f(t, \mathbf{v}_t, \mathbf{v}_{t-1}) = \frac{\Delta v}{\sqrt{1+\lambda t}}$ | $f(t, \mathbf{v}_t, \mathbf{v}_{t-1}) = \frac{1}{\sqrt{1+\lambda t \cdot \Delta t / \Delta v}}$ |
| Hyperbolic Decay | Decays very slowly, useful for long-term processes. | $f(t) = \frac{1}{1+\lambda t^2}$ | $f(t, \mathbf{v}_t, \mathbf{v}_{t-1}) = \frac{\Delta v}{1+\lambda t^2}$ | $f(t, \mathbf{v}_t, \mathbf{v}_{t-1}) = \frac{1}{1+\lambda t^2 \cdot \Delta t / \Delta v}$ |
| Logarithmic Decay | Provides a very slow decay, suitable for scenarios where recency fades gradually. | $f(t) = \frac{1}{\log(1+\lambda t+1)}$ | $f(t, \mathbf{v}_t, \mathbf{v}_{t-1}) = \frac{\Delta v}{\log(1+\lambda t+1)}$ | $f(t, \mathbf{v}_t, \mathbf{v}_{t-1}) = \frac{1}{\log(1+\lambda t \cdot \Delta t / \Delta v+1)}$ |
| Gaussian Decay | Decays rapidly at first and then flattens out, forming a bell-shaped curve. | $f(t) = e^{-\lambda t^2}$ | $f(t, \mathbf{v}_t, \mathbf{v}_{t-1}) = e^{-\lambda t^2} \cdot \Delta v$ | $f(t, \mathbf{v}_t, \mathbf{v}_{t-1}) = e^{-\lambda t^2 \cdot \Delta t / \Delta v}$ |

Fig. 3. Various Time Decay Functions

Compute decay as 1 / log(1 + (k * a * delta_t / cos_sim) + 1)

Gaussian Decay:

Compute decay as exp(-(k * a^2 * delta_t) / cos_sim)

Apply the computed decay to each corresponding embedding from the second onward

Return the adjusted embeddings

End Procedure

### 7.7 Diversity Analysis:

Diversity analysis was carried out with each models and with each decay functions.

Algorithm: Comprehensive Diversity Analysis with Decay, Cosine, and Time Adjustments

Inputs:

df - DataFrame containing user, timestamp, and embeddings data

model - Pre-trained embedding model

k - Decay constant for various decay calculations

Outputs:

Saves the computed embeddings with decay adjustments to numpy files

Procedure:

Initialize lists to store decay values for each decay function and user:

Basic decay functions: exp_dpe, inv_lin_dpe, inv_sqrt_dpe, hyperbolic_dpe, logarithmic_dpe, gaussian_dpe

Decay with cosine similarity: exp_cos_dpe, inv_lin_cos_dpe, inv_sqrt_cos_dpe, hyperbolic_cos_dpe, logarithmic_cos_dpe, gaussian_cos_dpe

Decay with cosine similarity and time: exp_cos_time_dpe, inv_lin_cos_time_dpe, inv_sqrt_cos_time_dpe, hyperbolic_cos_time_dpe, logarithmic_cos_time_dpe, gaussian_cos_time_dpe

Process each user's data:

Extract timestamps and embeddings for the user

Compute decay values for each decay type using embeddings and timestamps:

Apply basic decay functions

Apply decay functions adjusted by cosine similarity

Apply decay functions adjusted by cosine similarity and time differences

Append computed decay values to respective lists

Sum and Save Function:

Define a function to sum decay values for all users and save the results as numpy files

For each type of decay data list:

Sum across all users' decayed embeddings

Save the summed array to a numpy file

Diversity Analysis:

Calculate diversity metrics for each decay adjustment type using a custom diversity calculation function

Organize the diversity metrics into a DataFrame for analysis and visualization

Output:

Diversity metrics for each type of decay adjustment are saved and available for further analysis

End Procedure

**7.8 Application of Decay Functions:**

Various decay functions are applied to these embeddings to simulate the reduction in relevance of older data points. These functions include exponential, inverse linear, inverse square root, hyperbolic, logarithmic, and Gaussian. The decay functions are parametrized to adjust their impact, thereby allowing the examination of how sharply or gently they depreciate the older embeddings.

Algorithm: Plotting Decay Functions Over Time

Input:

timeline: Array of time indices from 1 to 100

k: Decay constant, example value 0.1

Output:

Plot of various decay functions over the specified timeline

Procedure:

Initialize the timeline from 1 to 100

Set the decay parameter k to 0.1

Calculate Decay Values:

Exponential Decay:

For each time t in timeline, compute exp(-k * t)

Inverse Linear Decay:

For each time t in timeline, compute 1 / (1 + k * t)

Inverse Square Root Decay:

For each time t in timeline, compute 1 / sqrt(1 + k * t)

Hyperbolic Decay:

For each time t in timeline, compute 1 / (1 + k * t^2)

Logarithmic Decay:

For each time t in timeline, compute 1 / log(1 + k * t + 1)

Gaussian Decay:

For each time t in timeline, compute exp(-k * t^2)

End Procedure

**7.9 Diversity Measurement:**

The diversity of the embeddings is measured using different similarities, assessing how similar or diverse the embeddings are from each other after the decay functions have been applied.

**Combined Diversity Matrix Across Models and Decay Functions**

**7.10 Model based performance matrix**

**7.11 Similarities based performance matrix**

# 8 COMPREHENSIVE ANALYSIS OF FINDINGS ACROSS DECAY FUNCTIONS, DIVERSITY METRICS, AND MODEL PERFORMANCE

**8.1 Overview** This analysis integrates various pre-trained models—MiniLM, DistilUSE Multilingual, MPNet, and Jina—to examine their performance across different decay functions and manage diversity metrics: basic, cosine, and cosine-time similarities.

**Analysis of Output Matrix and Graph**

8.1.1 **Exponential Decay (exp):** Shows moderate diversity in basic metrics, high diversity in cosine similarity, and excellent temporal consistency. This balance makes it suitable for environments where both recent and historical data influences are essential.

**8.1.2 Inverse Linear (inv_lin) and Square Root (inv_sqrt ):** These decays show a graduated reduction in diversity metrics, with inv_sqrt slightly outperforming inv_lin in handling outliers. This suggests their use in applications where historical data gradually diminishes in influence.

8.1.3 **Hyperbolic Decay:** Achieves the highest basic diversity, indicating its effectiveness in distinguishing between data points over a shorter term. However, its lower consistency over time may limit its application in long-term predictive modeling.

8.1.4 **Logarithmic Decay:** Similar to inverse functions but offers the slowest reduction in relevance, ideal for scenarios requiring long-term data retention without significant decay.

8.1.5 **Gaussian Decay:** Stands out with the highest cosine similarity, pointing to its efficiency in emphasizing recent data trends, making it highly suitable for real-time analytical applications.

8.2. **Summary of Model Performances Across Decay Functions**

• **MiniLM :** Consistently high performance in maintaining diversity across all metrics. Excels in cosine similarity across all decay functions, suggesting its effectiveness in capturing semantic relationships in multilingual contexts. Best under Gaussian decay, indicating robustness in adapting to recent data changes.

• **DistilUSE Multilingual:** Exhibits lower basic and cosine similarities compared to MiniLM, indicating a slight reduction in its ability to maintain diversity over time. Performs best under Gaussian decay for cosine similarity, aligning with MiniLM in terms of emphasizing recent data.

• **MPNet :** Shows a good balance across all diversity metrics, slightly outperforming DistilUSE in cosine similarity. This reflects its deep semantic understanding capabilities, especially under exponential and Gaussian decays. Best performance noted under Gaussian decay for both basic and cosine similarities.

• **Jina:** Registers the lowest diversity scores among the models, particularly in basic similarity, which might be due to its focus on similarity rather than diversity, as it is optimized for relevance-focused tasks like RAG models. Despite its lower scores, it shows a respectable performance in cosine similarity under Gaussian decay.

8.3 **Comparative Analysis and Diversity Metrics**

• **Cosine-Time Decay's Superiority:** Especially effective in capturing the maximum diversity across models, suggesting its utility in analyzing the evolution of data relationships over time.

• **Decay Function Effectiveness:** Gaussian decay emerges as the top performer, especially when recent data relevance is paramount. In contrast, logarithmic and hyper-bolic decays are better suited for applications that require a sustained influence of historical data.

TABLE 1
Diversity Matrix

| Model / Decay Function | Metric | MiniLM | DistilUSE Multilingual | MPNet | Jina |
|---|---|---|---|---|---|
| **Exponential** | Basic | 0.714106 | 0.659331 | 0.698469 | 0.386686 |
| | Cosine | 0.841320 | 0.786642 | 0.818740 | 0.729275 |
| | Cos-Time | 0.999453 | 0.999312 | 0.999429 | 0.999147 |
| **Inverse Linear** | Basic | 0.712126 | 0.656294 | 0.696391 | 0.379295 |
| | Cosine | 0.802726 | 0.754419 | 0.790047 | 0.568101 |
| | Cos-Time | 0.899301 | 0.872100 | 0.889009 | 0.802421 |
| **Inverse Square Root** | Basic | 0.712055 | 0.655778 | 0.696284 | 0.378312 |
| | Cosine | 0.843743 | 0.784945 | 0.818990 | 0.751743 |
| | Cos-Time | 0.999913 | 0.999873 | 0.999928 | 0.999874 |
| **Hyperbolic** | Basic | 0.723307 | 0.669123 | 0.707244 | 0.412020 |
| | Cosine | 0.817497 | 0.769464 | 0.803467 | 0.610245 |
| | Cos-Time | 0.924208 | 0.906596 | 0.920314 | 0.856247 |
| **Logarithmic** | Basic | 0.712133 | 0.655746 | 0.696314 | 0.378366 |
| | Cosine | 0.843750 | 0.784848 | 0.818969 | 0.750250 |
| | Cos-Time | 0.999924 | 0.999912 | 0.999927 | 0.999897 |
| **Gaussian** | Basic | 0.741066 | 0.688700 | 0.724950 | 0.462135 |
| | Cosine | 0.865656 | 0.814016 | 0.843493 | 0.794605 |
| | Cos-Time | 0.999503 | 0.999348 | 0.999477 | 0.999150 |

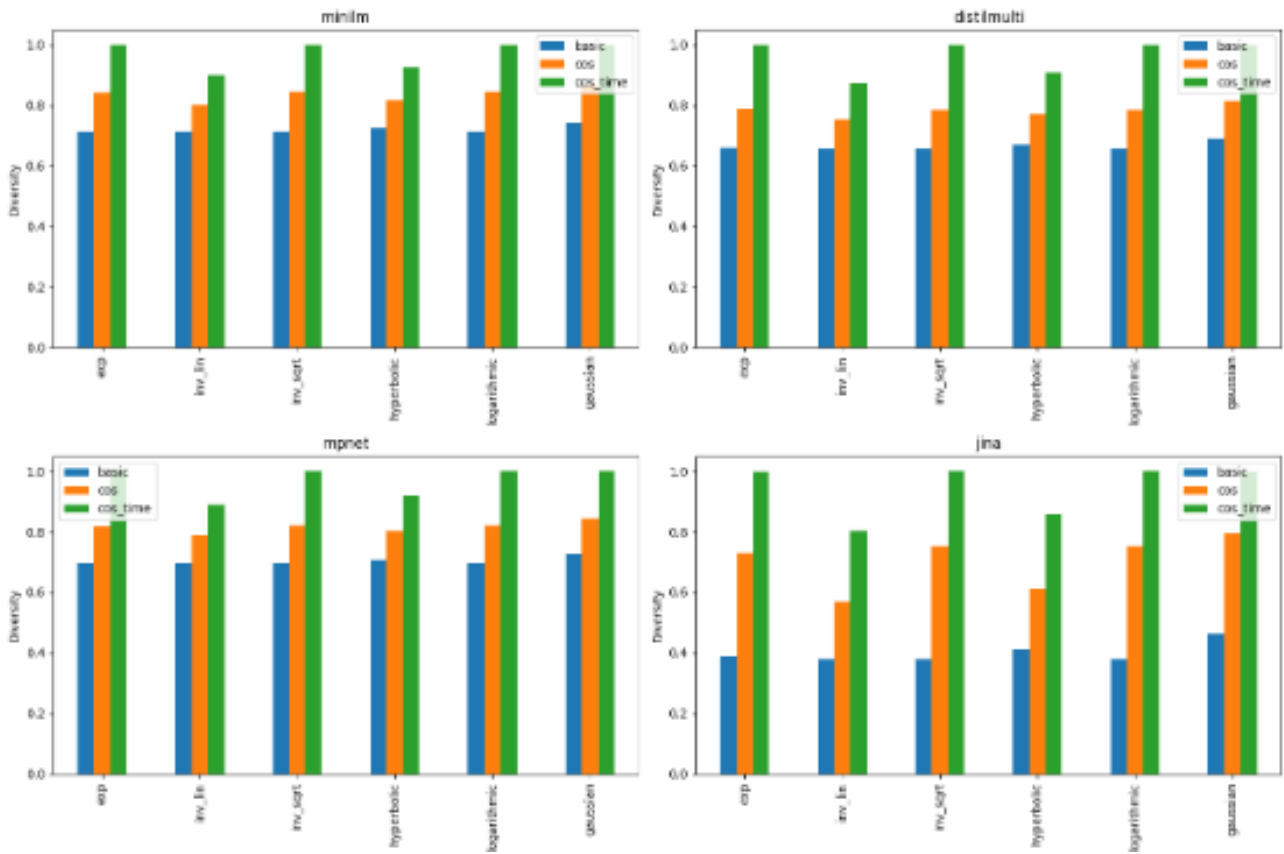

Fig. 4. Model based performance matrix

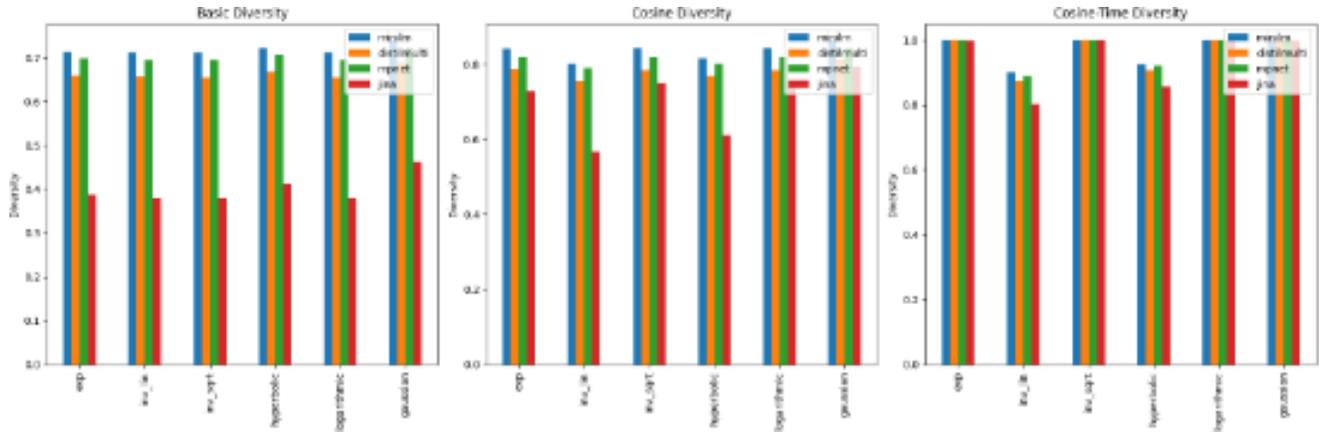

Fig. 5. Similarities based performance matrix

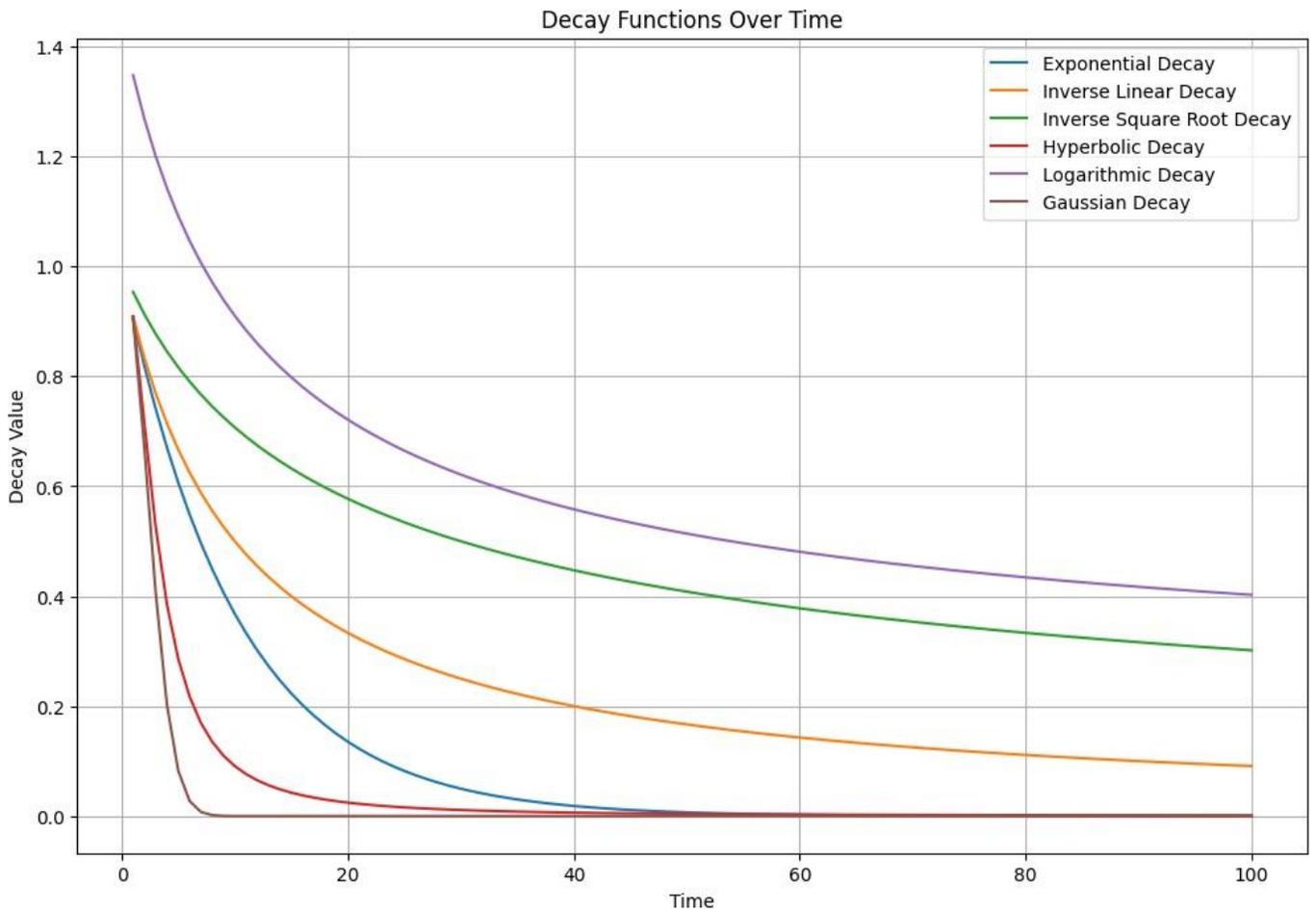

Fig. 6. – Decay Function over time

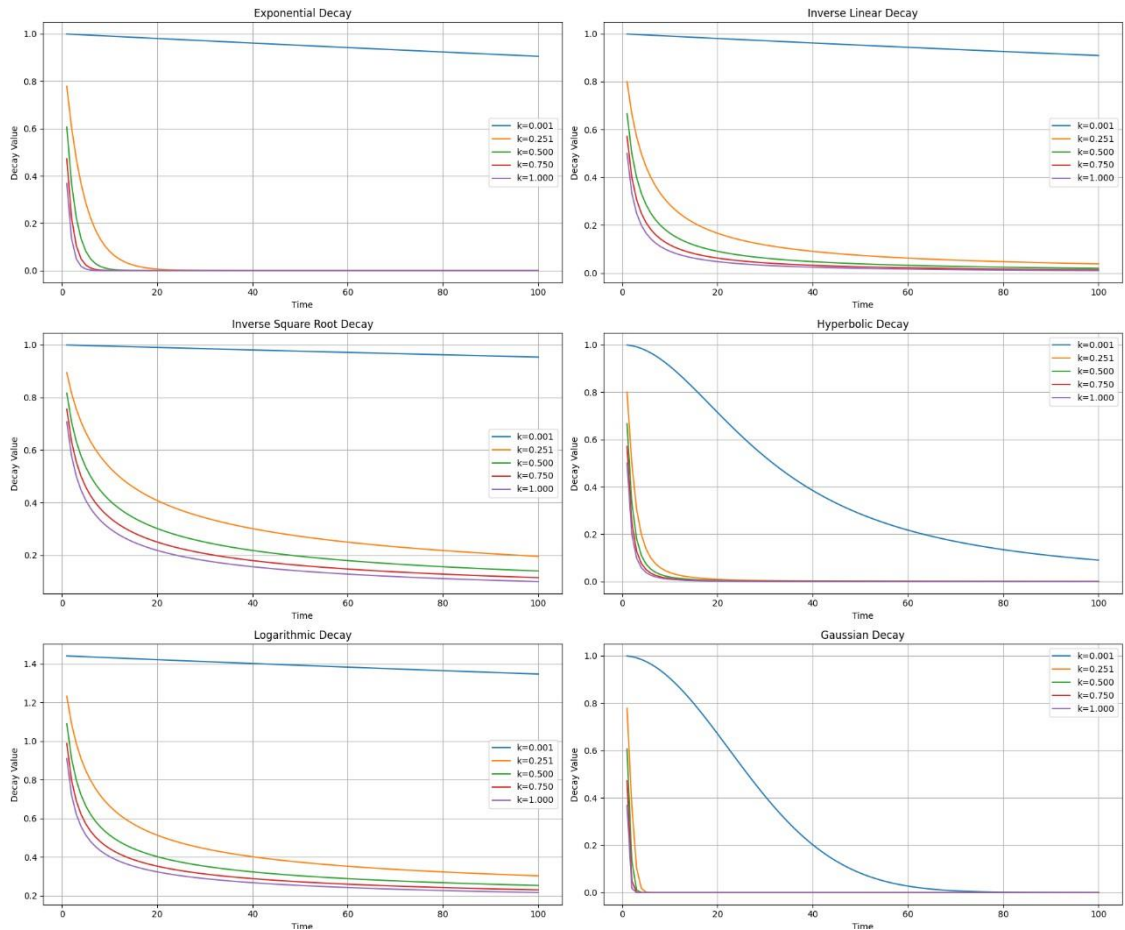

Fig. 7. Decay Function performance over time with various K Values

# 9 Supervised Evaluation Using Activity Data

## 9.1 Overview

Supervised Evaluation Using Activity Data aimed at analyzing User Activity Data to recommend activities based on user profiles and post embeddings. The first part involved time decay analysis with various decay functions and different models on User Data, including 100,100 individual user profiles and Tweet Data, resulting in approximately 20 million data points. In this second part, we analyze User Activity Data, such as likes, quote tweets, and retweets, by randomly selecting 30,000 instances, resulting in a total of 2,107,054 (2M) data points for liked tweets.

## 9.2 Unique Approach

The unique approach in this research involves several key elements that enhance the recommendation system:

9.2.1 **Using Activity Data:** By incorporating user activity data such as likes, quote tweets, and retweets, on top of user profile data and timeline data, the model gains an additional dimension to compute similarities. This comprehensive approach captures both static user characteristics and dynamic interactions, improving the relevance and accuracy of the recommendations.

9.2.1. **Pipeline Creation:** A well-structured pipeline was created for loading data, preprocessing, encoding using multiple embedding models, training, and evaluation. This pipeline ensures a systematic and reproducible process for analyzing and modeling the data.

9.2.2. **Multiple Embedding Models:** The use of multiple embedding models (MiniLM, DistilUse Multilingual, MPNET, and JINA) captures different aspects of the user activity data. Each model brings unique strengths, and their comparative analysis helps in identifying the best performer.

9.2.3. **Cosine Similarity Metric:** Utilizing cosine similarity as a metric for recommendation enhances the relevance of the suggestions by measuring the cosine of the angle between two non-zero vectors. This metric is particularly effective in high-dimensional spaces typical of embeddings.

9.2.4. **Neural Network Model with Normalization:** A neural network (ANN) model with normalization layers and mean squared error loss was developed. This model includes dense layers with ReLU activation functions to capture complex patterns in the data. The normalization step ensures that the input vectors are on the same scale, which is crucial for accurate similarity calculations.

9.3. **Explanation of the steps adopted:**

9.3.1. **Data Loading and Preprocessing :** The initial step

involves loading and preprocessing the Twitter activity data. This data comprises user interactions such as likes, quote tweets, and retweets. JSON files containing this activity data are loaded into a list called activity_data. For each entry in the activity_data, the text from each post is extracted and stored in a separate list called posts. This preprocessing step ensures that the data is structured and ready for further analysis.

9.3.2. **Model Setup and Encoding :** Once the data is pre-processed, the next step is to set up and initialize multiple embedding models using the Sentence Transformers library. Four models are used in this research: MiniLM, DistilUse Multilingual, MPNET, and JINA. These models are chosen for their varying capabilities in handling different languages and computational efficiencies. The text data in posts is then encoded into embeddings using each of these models. This step transforms the textual data into numerical vectors that can be used for similarity calculations and other machine learning tasks.

9.3.3. **Model Training and Evaluation :** To train and evaluate the models, a neural network (ANN) is created with input layers for the embeddings, followed by dense layers with ReLU activation functions. A custom cosine similarity function is used to measure the similarity between the embeddings. The model is compiled using the Adam optimizer and mean squared error loss. This ANN model helps in capturing the complex patterns within the embeddings and normalizes the input vectors to ensure accurate similarity calculations. The training involves loading .npy files containing the embeddings and fitting the model on these embeddings for evaluation.

9.3.4. **Pipeline Setup for Evaluating Similarity :** The approach involved setting up a pipeline to process the results obtained from the time decay analysis conducted on various models using user data and their tweet timelines. This pipeline was designed to incorporate activity data, such as likes, quote tweets, and retweets, to further evaluate the similarity between user profiles and their interactions. By systematically processing the embeddings through this pipeline, the additional activity data provided an enhanced dimension for accurately assessing similarities and making more precise activity recommendations.

9.3.5. **Result Compilation :** The results are compiled into a DataFrame for further analysis. This involves iterating through the directory containing the .npy files, training the model on each file, and storing the evaluation results in a dictionary. These results are then transformed into a DataFrame format, which facilitates easy comparison and visualization.

9.3.6. **Algorithm: Neural Network Model Creation and Training**

**Input:**
e1, e2, e3, e4: Encoded embeddings for each post using four different models

directory: Directory path containing .npy files

**Output:**
Results: Dictionary containing evaluation results

**Procedure:**
Import necessary modules from tensorflow.keras.
Define function create_model(): a. Initialize input layers input1 and input2 with shape (512,). b. Define dense layers

dense1 and dense2 with ReLU activation. c. Define custom cosine similarity function:

Normalize input x and y.

Return sum of element-wise multiplication of x and y.
d. Compile model with inputs [input1, input2] and output cos_sim, optimizer 'adam', loss 'mean_squared_error'. e. Return model.

Define function train_evaluate(file, e): a. Load input_1 from file. b. Set input_2 as e. c. Initialize model using create_model(). d. Fit model on input_1 and input_2 for 1 epoch. e. Return mean of sigmoid predictions on input_1 and input_2.

Initialize an empty dictionary results.
Evaluate Models: a. For each file in directory:
If file ends with '.npy':
If file starts with '1':
Store result of train_evaluate(file, e1) in results[file].
Else if file starts with '2':
Store result of train_evaluate(file, e2) in results[file].
Else if file starts with '3':
Store result of train_evaluate(file, e3) in results[file].
Else if file starts with '4':
Store result of train_evaluate(file, e4) in results[file].
End Procedure

**9.3.7. Combined Supervised Evaluation matrix Using Activity Data:**

This table summarizes the evaluation results for each model (MiniLM, DistilUse, MPNET, JINA) across different decay functions (Gaussian, Inv_Sqrt, Logarithmic, Exp, Inv_Lin, Hyperbolic) and metrics (Basic, Cos, Cos_Time).

**9.3.8. Findings from the Combined Supervised Evaluation Using Activity Data**

9.3.8.1 **Overall Performance:** The Jina model consistently outperformed the other models across most metrics and decay functions, achieving the highest scores in both Cosine and Cos-Time metrics. Jina performs the best for recommending activities accurately, effectively matching the profile embeddings to the post embeddings.

9.3.8.2 **Model Ranking:** The overall ranking of models based on their performance is: Jina > DistilUSE Multilingual > MPNet > MiniLM.

9.3.8.3 **Gaussian Decay Function:** For the Gaussian decay function, Jina achieved a perfect score in the Cos-Time metric (1.000000), indicating its superior ability to capture activity data dynamics. DistilUSE Multilingual and MiniLM also performed well, but Jina was the clear leader.

9.3.8.4 **Inverse Square Root Decay Function:** Jina again performed exceptionally well, with perfect scores in the Cos-Time metric. DistilUSE Multilingual followed closely, while MiniLM and MPNet showed lower performance compared to Jina.

9.3.8.5 **Logarithmic Decay Function:** The Jina model led in the Cos-Time metric with a score of 0.965639. DistilUSE Multilingual and MPNet showed good performance, but MiniLM lagged behind.

9.3.8.6 **Exponential Decay Function:** Jina scored highest in the Cos-Time metric, followed by DistilUSE Multilingual and MPNet. MiniLM had the lowest scores among the models for this decay function.

9.3.8.7 **Inverse Linear Decay Function:** Jina maintained the lead with the highest scores across all metrics, partic-ularly in the Cos-Time metric. DistilUSE Multilingual and MPNet were moderately successful, while MiniLM showed comparatively lower performance.

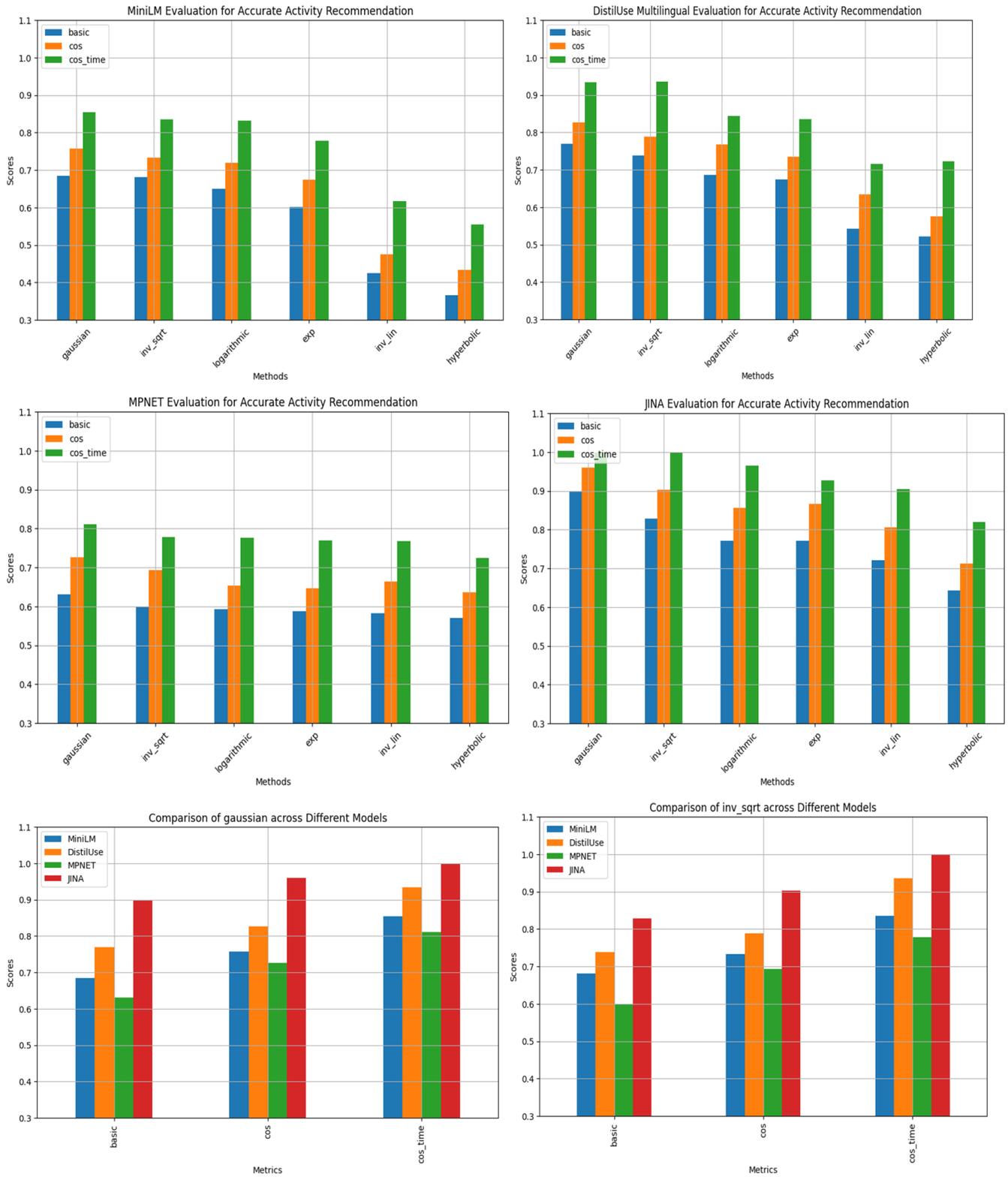

Fig. 8. Model wise performance

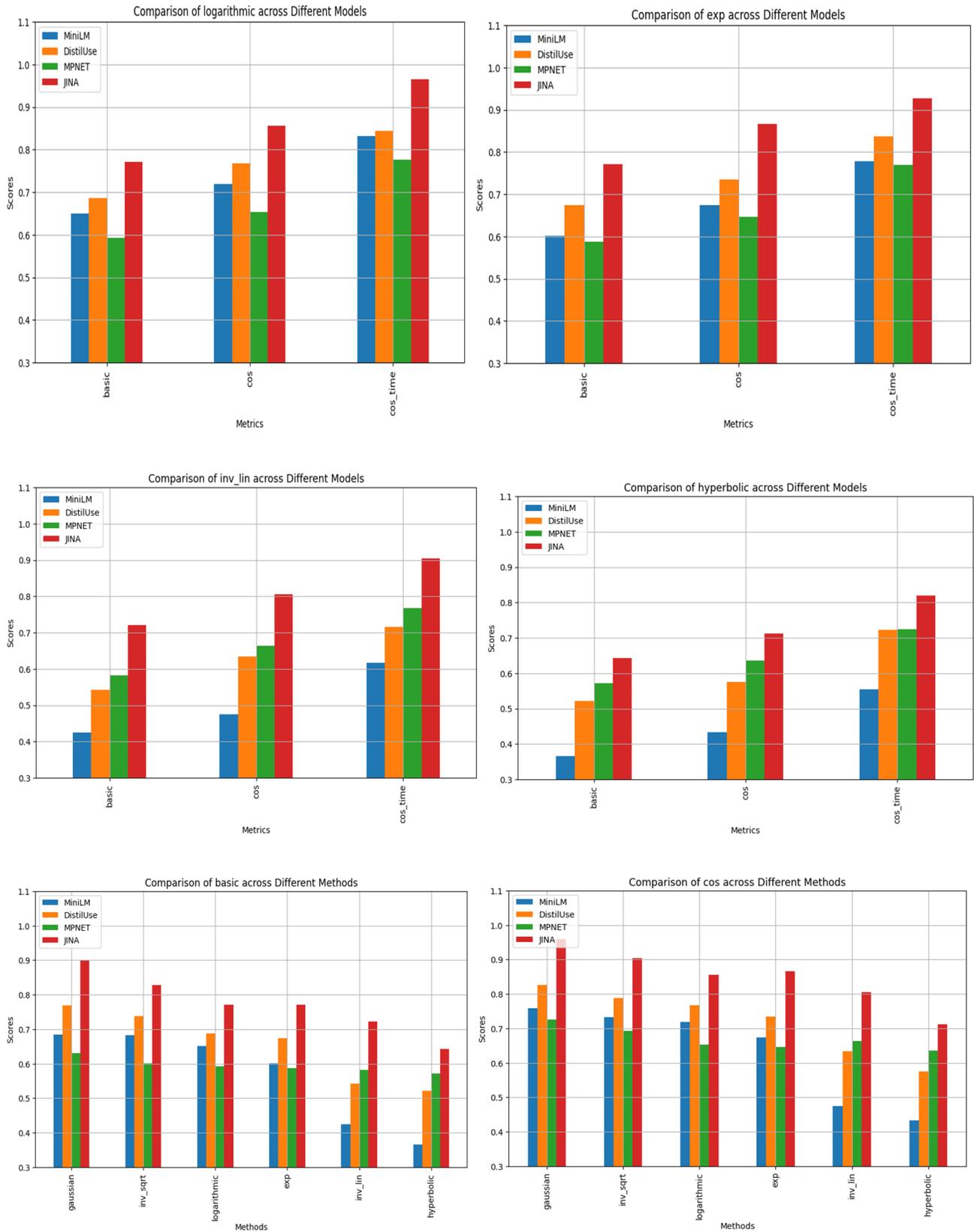

Fig. 9. Combined analysis

TABLE 2
Evaluation matrix Using Activity Data

| Model / Decay Function | Metric | MiniLM | DistilUSE Multilingual | MPNet | Jina |
|---|---|---|---|---|---|
| Gaussian | Basic | 0.684799 | 0.768862 | 0.631093 | 0.899110 |
| | Cosine | 0.758124 | 0.826680 | 0.726407 | 0.960034 |
| | Cos-Time | 0.854276 | 0.933938 | 0.811347 | 1.000000 |
| Inverse Square Root | Basic | 0.681950 | 0.737953 | 0.600583 | 0.827555 |
| | Cosine | 0.733301 | 0.788345 | 0.692652 | 0.903512 |
| | Cos-Time | 0.835079 | 0.936106 | 0.777927 | 1.000000 |
| Logarithmic | Basic | 0.650677 | 0.687051 | 0.592280 | 0.771490 |
| | Cosine | 0.719356 | 0.767229 | 0.653591 | 0.856169 |
| | Cos-Time | 0.831182 | 0.843845 | 0.775865 | 0.965639 |
| Exponential | Basic | 0.601358 | 0.673745 | 0.587857 | 0.770413 |
| | Cosine | 0.673760 | 0.734020 | 0.645904 | 0.866156 |
| | Cos-Time | 0.777315 | 0.836165 | 0.769279 | 0.927258 |
| Inverse Linear | Basic | 0.424232 | 0.543239 | 0.582079 | 0.721708 |
| | Cosine | 0.474385 | 0.634266 | 0.663415 | 0.805891 |
| | Cos-Time | 0.616621 | 0.715388 | 0.767864 | 0.905271 |
| Hyperbolic | Basic | 0.366394 | 0.522032 | 0.571216 | 0.642350 |
| | Cosine | 0.433326 | 0.575382 | 0.636290 | 0.711610 |
| | Cos-Time | 0.555031 | 0.721971 | 0.724740 | 0.819516 |

9.3.8.8 **Hyperbolic Decay Function:** Jina's performance was strong, especially in the Cos-Time metric, demonstrating its robustness across various decay functions. DistilUSE Multilingual and MPNet had lower scores, with MiniLM showing the least performance.

9.3.8.9 **Metric Comparison:** The ranking of metrics based on their effectiveness in capturing activity data is: Cos-Time > Cosine > Basic. Cos-Time consistently showed the highest scores, highlighting its effectiveness in evaluating activity data.

9.3.8.10 **Decay Function Comparison:** The ranking of decay functions based on model performance is: Gaussian > Inverse Square Root > Logarithmic > Exponential > Inverse Linear > Hyperbolic. This indicates that the Gaussian decay function is the most effective in this context.

## 10 COMPARISON OF DYNAMIC EMBEDDINGS WITH STATIC EMBEDDINGS

### 10.1 Approach

To carry out Static embeddings, the primary dataset consists of Twitter activity data across multiple JSON files. The approach involves setting up the environment with necessary libraries like sentence-transformers and tqdm for progress tracking. Multiple pre-trained models from the sentence-transformers library are employed, specifically all-MiniLM-L6-v2, distiluse-base-multilingual-cased-v2, all-mpnet-base-v2, and jina-embeddings-v2-base-en. These models are loaded onto a CUDA device for efficient processing. Additionally, profile data is read from a CSV file containing user information and tweets.

### 10.2 Process

The process involves several steps:

*10.2.1* **Data Loading:** JSON files containing Twitter activity timelines are loaded and parsed.

*10.2.2* **Text Extraction:** Tweets are extracted from the parsed JSON data and stored in a list.

*10.2.3* **Embedding Generation:** Each pre-trained model is used to generate embeddings for the extracted tweets.

*10.2.4* **Diversity Calculation:** A function calculates the diversity of the embeddings using cosine similarity. This diversity metric helps in comparing the uniqueness of the embeddings generated by different models.

*10.2.5* **Static Diversity Comparison:** A bar plot visualizes the diversity scores of the static embeddings generated by each model.

### 10.3 Diversity results of Static Embeddings

For diversity, the maximum value for static embeddings was achieved by the mpnet model at 0.36. This value is significantly lower compared to the diversities of dynamic embeddings. The primary reason for this disparity is the method used to calculate static embeddings: averaging all the tweets rather than applying a decay function. This averaging process substantially reduces the uniqueness of each profile, thereby lowering the overall diversity.

### 10.4 Accuracy calculation for recommendation by static embeddings

The Jina static embeddings performed the best for accurately identifying recommendations among the static models. However, they still fall short when compared to dynamic embeddings. The aggregated static embeddings fail to effectively match the user profiles to their activities, highlighting the limitations of static methods in capturing user behavior accurately.

### 10.5 Findings: Static vs Dynamic Embeddings

The graph illustrates that dynamic embeddings generally exhibit higher diversity scores compared to static embeddings across all models. This supports the observation that dynamic embeddings are better at capturing a broader and more up-to-date range of user behaviors and preferences.

This visualization clearly shows that dynamic embeddings significantly outperform static embeddings in accuracy across all models. This supports the conclusion that dynamic embeddings are more effective in applications requiring precise and up-to-date personalization, such as recommendation systems on social media platforms.

TABLE 3
Diversity result of static embedding

| Model | Diversity |
|-------|-----------|
| MiniLM | 0.2384185791015625 |
| Multi | 0.27881393432617185 |
| mpnet | 0.361625671386717 |
| jina | 0.1920928955078124 |

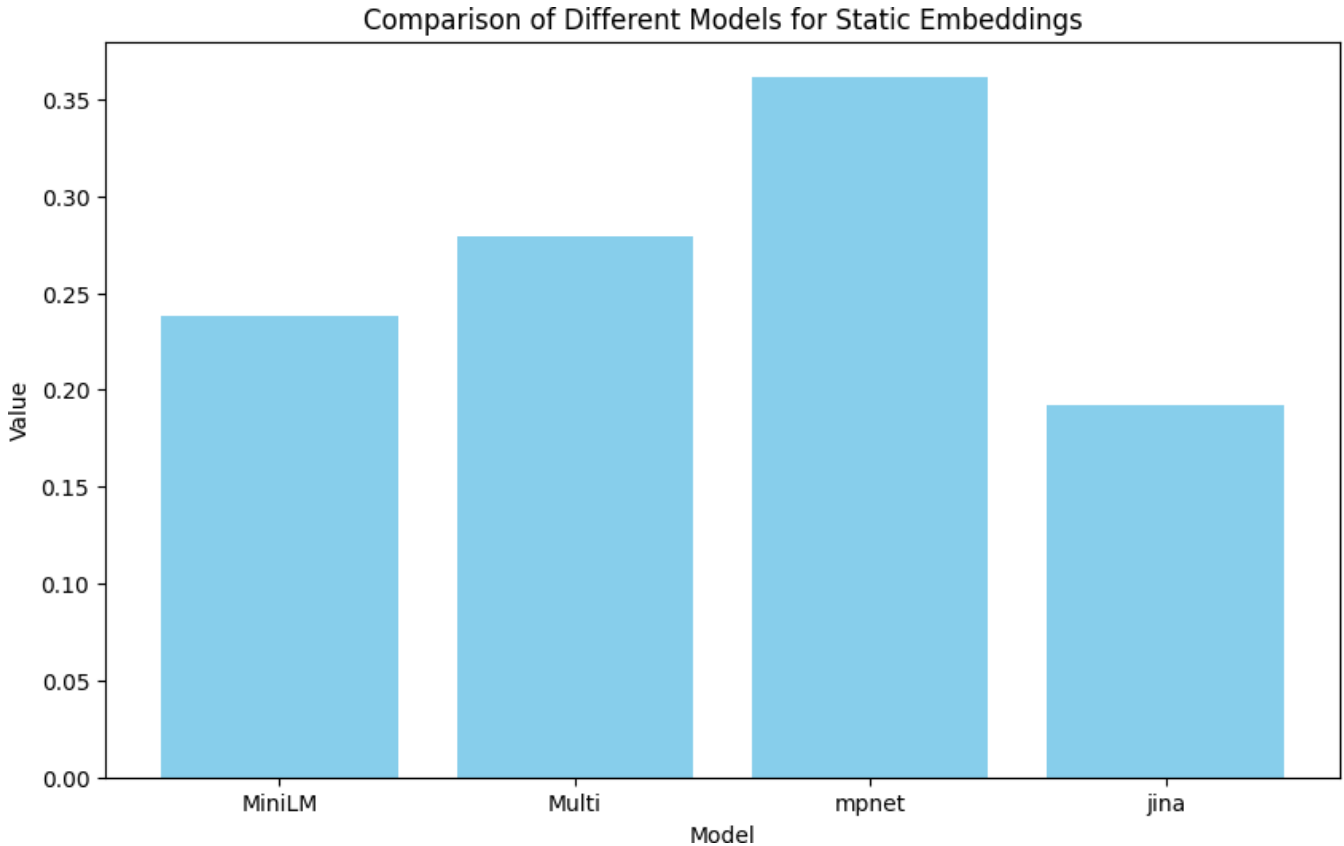

Fig. 10. Comparison of Diversity result of Different models

TABLE 4
Recommendation accuracy of various models

| Model | Accuracy |
|-------|----------|
| MiniLM | 0.01782471 |
| Multi | 0.25269645 |
| mpnet | 0.31811421 |
| jina | 0.39415552 |

### 10.5.1 Adaptability and Relevance:

• **Dynamic embeddings** demonstrate high adaptability with scores approaching 1.0 in Cos-Time metrics (e.g., MiniLM: 0.9995, MPNet: 0.9999), showing their ability to stay current with evolving user profiles.

  • **Static embeddings**, in contrast, maintain a consistent diversity measurement (e.g., MiniLM: 0.238, MPNet: 0.362) which does not reflect changes over time.

### 10.5.2 Diversity and Accuracy:

**Dynamic embeddings** often achieve higher diversity scores (e.g., MPNet in Cosine metric: 0.843 under Gaussian decay) compared to static embeddings, indicating a better capture of a broad range of user behaviors.

**Accuracy** of dynamic embeddings in real-time applications also tends to be higher. For example, Jina shows a Cos-Time metric score of 1.000 in the Gaussian decay setting, highlighting its precision in capturing user preferences dynamically.

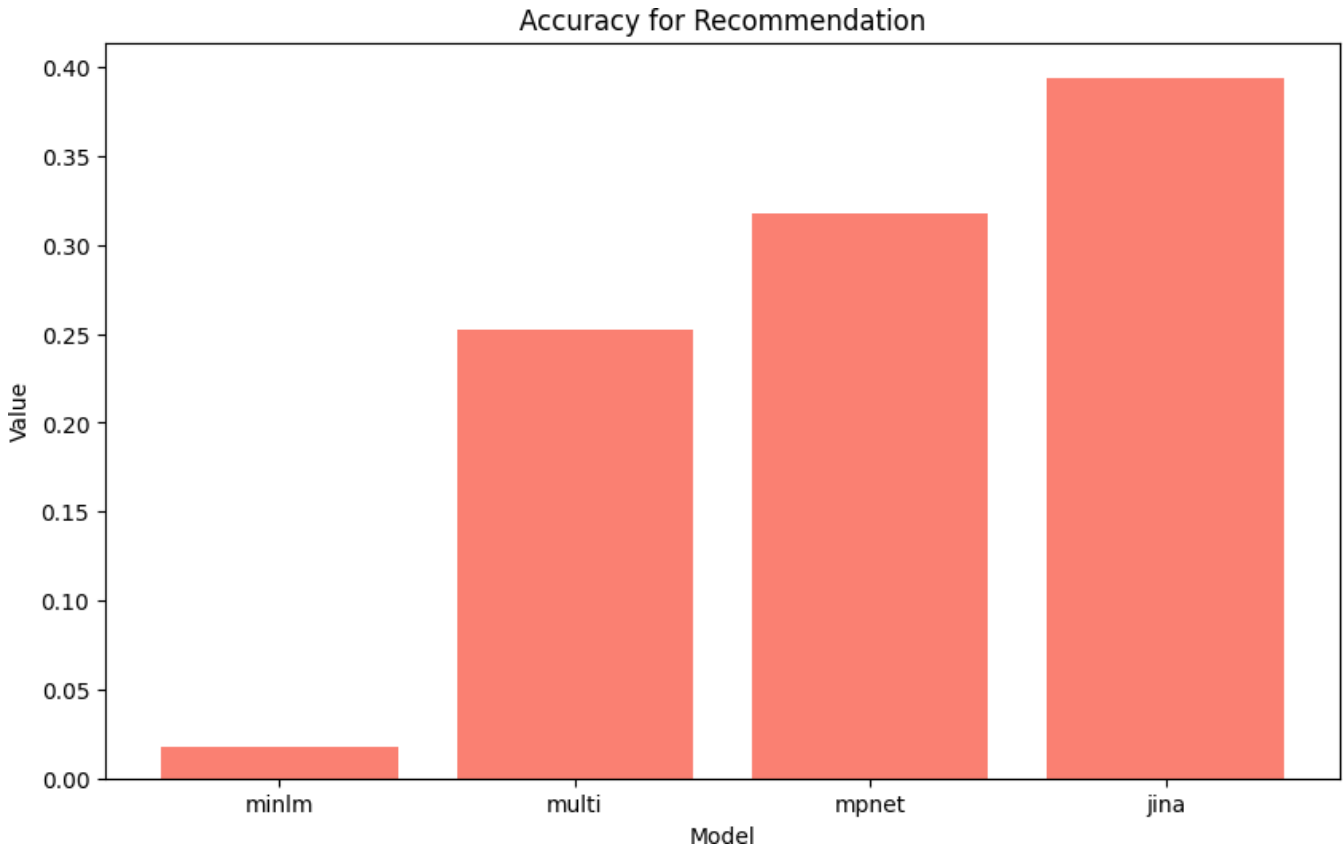

Fig. 11. Recommendation accuracy of various models

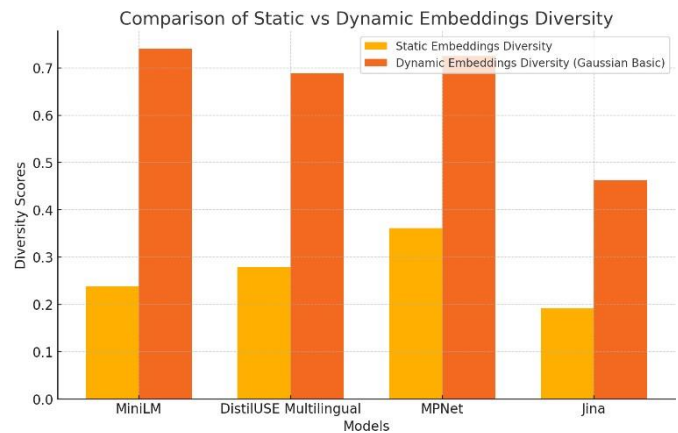

Fig. 12. Comparison ofStatic vs Dynamic Embeddings Diversity

### 10.5.3  Performance Across Decay Functions:

**Exponential and Gaussian decays** are especially effective in prioritizing recent data. Gaussian decay shows high scores across all models in the Cos-Time metric (e.g., Jina: 1.000, MiniLM: 0.9995).

10.5.4. **Logarithmic and hyperbolic decays** maintain relevance of older data better, with slower reduction rates (e.g., Logarithmic Cos-Time for MiniLM: 0.9999), useful in applications needing long-term behavioral analysis.

**Suitability for Real-Time Applications**:

10.5.5. Dynamic embeddings using **Gaussian decay** are notably suited for real-time applications, shown by high Cos-Time scores, indicating they adjust quickly to new user data (Jina: 1.000, DistilUSE Multilingual: 0.933).

### 10.5.6. **Computational Efficiency**:

While dynamic embeddings offer updated insights, they require ongoing computation as new data arrives, contrasting with static embeddings which are computed once (e.g., static diversity for MiniLM: 0.238 vs dynamic diversity under Gaussian: 0.741).

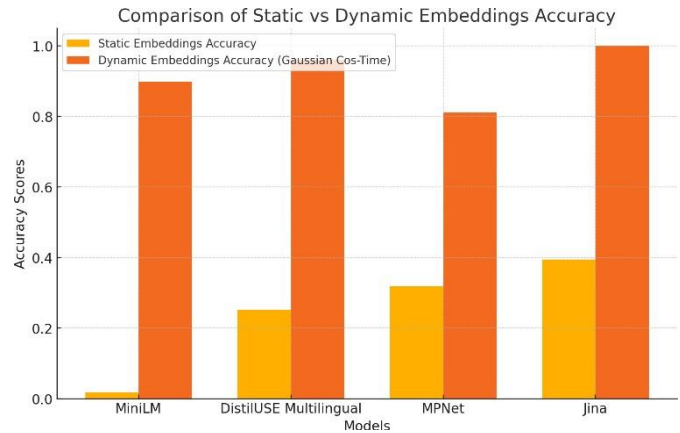

Fig. 13. Comparison of Static vs Dynamic Embeddings Accuracy

### 10.5.4 Application Specificity:

Dynamic embeddings are recommended for environments with frequent user interaction changes, as shown by their superior performance in dynamic settings (e.g., Jina's consistent high scores in dynamic settings).

Static embeddings may still be useful in stable, less change-intensive environments, offering a simpler computational approach without the need for frequent updates.

## 11 INTERPRETATION OF THE RESULTS IN THE CONTEXT OF THE RESEARCH QUESTION

The study explores the efficacy of different decay functions and embedding models in capturing user behavior dynamics for recommendation systems. The findings suggest that Gaussian and exponential decay functions, when combined with models like Jina, MiniLM and MPNet, effectively prioritize recent interactions while still maintaining historical context, thereby enhancing recommendation systems' accuracy and relevance. This addresses the research question concerning how different temporal decay functions influence the adaptability of dynamic embeddings in social media environments.

## 12 COMPARISON WITH PREVIOUS STUDIES

### 12.1 Decay Function Effectiveness:

• **Similarity with Existing Work**: The study confirms the effectiveness of Gaussian decay in capturing recent data trends, which is consistent with the findings of Covington et al. (2016) [23] in their study on deep neural networks for YouTube recommendations.

• **Differences**: The study also highlights the importance of logarithmic and hyperbolic decays in maintaining longer historical contexts, which is not explicitly discussed in Covington et al. (2016). [23]

### 12.2 Model Performance:

**Similarity with Existing Work**: The study shows that MiniLM consistently performs well across all metrics, which is consistent with the findings of Chen et al. (2018) in their survey on session-based recommender systems. [22]

**Differences**: The study also highlights the strengths of MPNet in balancing diversity metrics and its ability to adapt to recent data changes, which is not explicitly discussed in Chen et al. (2018).

### 12.3 Diversity Metrics:

**Similarity with Existing Work**: The study emphasizes the importance of cosine similarity in capturing semantic relationships, which is consistent with the findings of He and McAuley (2016) [25] in their study on modeling the visual evolution of fashion trends with one-class collaborative filtering.

**Differences**: The study also highlights the effectiveness of cosine-time decay in capturing the maximum diversity across models, which is not explicitly discussed in He and McAuley (2016). [25]

## 13 RECOMMENDATIONS:

13.1. **Similarity with Existing Work**: The study recommends a balanced approach to decay functions and model selection based on application requirements. This aligns with the findings of Sarwar et al. (2001)[32] in their study on item-based collaborative filtering recommendation algorithms.

13.2. **Differences**: The study provides more specific recommendations for real-time systems and historical data analysis, which are not explicitly discussed in Sarwar et al. (2001). [32]

## 14 REAL-WORLD DATA:

The study's use of real-world data from Twitter, including user profiles and tweet activity, provides a more practical and applicable context for the evaluation of recommender systems. This aligns with the focus on real-world data in other studies, such as Archak (2010) [24], which analyzed strategic behavior in crowdsourcing contests on TopCoder.com.

## 15 SUGGESTIONS FOR FUTURE RESEARCH

• Exploring Additional Models: Future research could explore the integration of newer or less common machine learning models that might offer improved performance or efficiency.

• Cross-platform Validation: Validating the findings across different social media platforms or even in different contexts (like e-commerce or content streaming services) could help in understanding the broader applicability of the proposed methods.

• Real-time Implementation: Investigating the real-time implementation of these models and decay functions in live environments could provide insights into practical challenges and performance issues.

• User Feedback Incorporation: Incorporating user feedback mechanisms to dynamically adjust the decay rates and model parameters could enhance the adaptability and accuracy of recommender systems.

• Ethical and Privacy Considerations: Further research could also explore the ethical implications and privacy concerns related to the use of dynamic embeddings in user profiling.

## 16 SUMMARY OF KEY FINDINGS

• Decay Function Effectiveness : The study validates the effectiveness of Gaussian decay for emphasizing recent interactions ,aligning with existing literature that underscores the importance of recency in recommender systems. Also it highlights the unique value of logarithmic and hyperbolic decays in keeping historical context, offering a more nuanced approach to temporal data analysis in dynamic embeddings.

• Model Performance: From the various models tested, Jina, MiniLM and MPNet demonstrated robust performance, excelling in balancing recent and historical data influences. This suggests their suitability for complex recommendation tasks that require a nuanced understanding of user behaviors over time..

• Diversity Metrics : The use of cosine and cosine-time similarity metrics were particularly effective in capturing semantic relationships and maximizing diversity across models, pointing towards their potential in enhancing the accuracy and relevance of recommendations. . .

## 17 FINAL REMARKS ON THE STUDY'S CONTRIBUTIONS

This research contributes significantly to the field of recommender systems by offering a comprehensive evaluation of how different decay functions affect the performance of dynamic embeddings. It bridges a gap in existing research by systematically analyzing the impact of less commonly used decay functions and by exploring their implications in real-world social media data. The study's methodological rigor and detailed analysis provide valuable benchmarks for future research and practical applications in dynamic user profiling.

## 18 POTENTIAL APPLICATIONS OF THE RESEARCH

• Social Media Platforms: Adopting the approach and the findings of the study can help various social-media platforms including multilingual platforms to refine their content recommendation algorithms for better alignments with user preferences, enhancing user engagement and satisfaction..

• E-commerce: Online retailers, e-shopping portals can utilize the approach and the findings of the study to improve their product recommendation systems, also potentially increasing conversion rates by presenting products that align more with the evolving interests of their customers. Also this can be help ful for online advertisement planning based on the interest and relevance of the consumers.

• Content Streaming Services: OTT platforms like Netflix or Spotify, applying dynamic embeddings with appropriate decay functions can improve content recommendations ,keeping them fresh and relevant to user choice that change over time.

• Customer Relationship Management (CRM): CRMs can integrate these findings to better track customer interactions over time, providing insights that help in customising marketing strategies and customer service initiatives as per the customer behaviour trends.

• Adaptive Learning Systems: In educational technology, adapting content delivery based on dynamic user profile embeddings can personalize learning experience ,also it can improve outcomes by aligning educational content with the changing needs and proficiency levels of learners.